\documentclass[12pt]{article}
\usepackage[rm,bf,raggedright]{titlesec}    
\usepackage{times}
\usepackage[tbtags]{amsmath}
\usepackage{amsfonts}
\usepackage[amssymb,pstricks]{SIunits}
\usepackage[figuresright]{rotating}         
\usepackage{diagnose}                       
\usepackage{graphicx}                      
\usepackage{lineno}                         
\usepackage{lscape}                         
\usepackage{longtable}                      
\usepackage{threeparttable}                 
\usepackage{natbib}                         
\usepackage{booktabs}                       
\usepackage{cases}
\usepackage{textcomp}
\usepackage{dcolumn}                        
\usepackage{warpcol}                        
\usepackage{placeins}
\usepackage[printonlyused]{acronym}         
\usepackage{mhequ}
\usepackage[letterpaper,includehead,height=9in]{geometry}
\usepackage{extramarks}
\usepackage{fancyhdr}                       
\usepackage{nicefrac}                       
\usepackage{endnotes}                       
\usepackage{float}                          
\usepackage{tabularx}                       
\usepackage{ltabptch}
\usepackage{warpcol}
\usepackage{url}                            
\usepackage{ccaption}                       
\usepackage[version=3]{mhchem}              
\usepackage{bm}                             
\usepackage{icomma}                         
\usepackage{setspace}                       
\usepackage{ragged2e}                       
\usepackage{tabularx}                       
\addunit{\week}{wk}%
\addunit{\mont}{mo}%
\addunit{\yr}{yr}%
\addunit{\microgramperliterrp}{\upmu\gram\usk\reciprocal\liter}
\addunit{\milligramperliterrp}{\milli\gram\usk\reciprocal\liter}
\addunit{\gramperliterrp}{\gram\usk\reciprocal\liter}
\addunit{\milligrampercubicmetre}{\milli\gram\usk\rpcubic\metre}
\addunit{\grampercubicmetrerp}{\gram\usk\rpcubic\metre}
\addunit{\milligramperliterkilogramrp}{\milli\gram\usk\reciprocal\liter\usk\reciprocal\kilogram}
\addunit{\kilometre}{\kilo\metre} \addunit{\centimetre}{\centi\metre}
\addunit{\metreperhournp}{\metre\usk\reciprocal\hour}
\addunit{\metreperdaynp}{\metre\usk\reciprocal\dday}
\addunit{\centimetrepersecondnp}{\centimetre\usk\reciprocal\second}
\addunit{\cubicmetreperhourrp}{\cubicmetre\usk\reciprocal\hour}
\addunit{\cubiccentimetrepersecondrp}{\cubiccentimetre\usk\reciprocal\second}
\addunit{\cubiccentimetreperhourrp}{\cubiccentimetre\usk\reciprocal\hour}
\addunit{\squaremetreperhourrp}{\squaremetre\usk\reciprocal\hour}
\addunit{\milligramperdayrp}{\milli\gram\usk\reciprocal\dday}
\addunit{\gramperkilogramrp}{\gram\usk\reciprocal\kilogram}
\addunit{\milligramperkilogramrp}{\milli\gram\usk\reciprocal\kilogram}
\addunit{\microgramperkilogramperdayrp}{\upmu\gram\usk\reciprocal\kilogram\usk\reciprocal\dday}
\addunit{\milligramperkilogramperdayrp}{\milli\gram\usk\reciprocal\kilogram\usk\reciprocal\dday}
\addunit{\cubicmilliliterpergramrp}{\milli\cubic\liter\usk\reciprocal\gram}
\addunit{\grampercubiccentimetrerp}{\gram\usk\rpcubic\centimetre}
\addunit{\becpercubicmetrerp}{\becquerel\usk\rpcubic\metre}
\addunit{\becsecondpercubicmetrenp}{\becquerel\usk\reciprocal\second\usk\rpcubic\metre}
\addunit{\bechourpercubicmetrerp}{\becquerel\usk\hour\usk\rpcubic\metre}
\addunit{\becyearpercubicmetrenp}{\becquerel\usk\yocto\usk\rpcubic\metre}
\addunit{\picocurieliterrp}{\pico\curie\usk\reciprocal\liter}
\addunit{\joulesecondpercubicmetrenp}{\joule\usk\second\usk\rpcubic\metre}
\addunit{\joulehourpercubicmetrenp}{\joule\usk\hour\usk\rpcubic\metre}
\addunit{\worklevel}{WL} \addunit{\worklevelmonth}{WLM}
\addunit{\millivoltsperliter}{\mega\electronvolt\usk\reciprocal\liter}
\addunit{\millivoltspercubicmetrenp}{\mega\electronvolt\usk\rpcubic\metre}
\addunit{\peryear}{\usk\reciprocal\yr}
\addunit{\millisievertperyear}{\milli\sievert\peryear}
\renewcommand\descriptionlabel[1]%
{\hspace{\labelsep}\textrm{#1}}
\newdimen\biblioindent
\newenvironment{notation}[1]%
  {\ifx#1\item\ClassError{article}{%
    The notation environment MUST have an argument:\MessageBreak
    the longest symbol to appear in the listing}%
    {Stop and add an argument to \protect\begin{notation}{sym}}\fi
   \section*{\sc Notation}\begin{list}{}{%
     \settowidth{\labelwidth}{#1}%
     \setlength{\itemsep}{0.0in}\setlength{\parsep}{0.0in}%
     \setlength{\leftmargin}{\labelwidth}%
     \setlength{\labelsep}{1em}%
     \addtolength{\leftmargin}{\labelsep}}}{\end{list}}
\begin{document}
%
\renewcommand{\footnote}{\endnote}
\renewcommand{\enotesize}{\normalsize}
%
\captiondelim{. }%
\captionstyle{\raggedright}%
\captionnamefont{\bfseries}%
\captiontitlefont{\bfseries}%
%
\pagestyle{fancy}%
\lhead{\sl P.\@ Zhu}%
\fancyhead[R]{\thepage}%
\fancyfoot{}
%
\titlelabel{}
\titleformat*{\section}{\large\scshape\Centering}
\titleformat*{\subsection}{\scshape}
\titleformat*{\subsubsection}{\itshape}
%
%
\begin{titlepage}
\title{Hydrological and tectonic strain forces measured from a karstic cave using extensometers}
\author{Ping Zhu \footnote{zhuping@oma.be Royal Observatory of Belgium, ORB-AVENUE CIRCULAIRE 3, 1180, Bruxelles, Belgium.} Michel van Ruymbeke \footnote{labvruy@oma.be Royal Observatory of
Belgium, ORB-AVENUE CIRCULAIRE 3, 1180, Bruxelles, Belgium.} Yves
Quinif \footnote{Yves.Quinif@fpms.ac.be Institut Jules Cornet
(G\'{e}ologie),Facult\'{e} Polytechniqude de Mons, Rue de Houdain 9,
7000 Mons,Belgium} Thierry Camelbeeck
\footnote{Thierry.Camelbeeck@oma.be Royal Observatory of Belgium,
ORB-AVENUE CIRCULAIRE 3, 1180, Bruxelles, Belgium.} Philippe Meus
\footnote{philippe.meus@spw.wallonie.be Service Public de Wallonie
Direction g\'{e}n\'{e}rale op\'{e}rationnelle Agriculture,
Ressources naturelles et Environnement D\'{e}partement de
l'Environnement et de l'Eau Direction des Eaux souterraines 15
Avenue Prince de Li\'{e}ge, B-5100 Jambes, Belgium}}

\date{\today}
\maketitle
\end{titlepage}
\clearpage
%
%
In order to monitor the hydrological strain forces of the karst
micro fissure networks and local fault activities, six capacitive
extensometers were installed inside a karstic cave near the
midi-fault in Belgium. From 2004 to 2008, the nearby Lomme River
experienced several heavy rains, leading to flooding inside the
Rochefort cave. The highest water level rose more than thirteen
meters, the karstic fissure networks were filled with water, which
altered the pore pressure of the cave. The strain response to the
hydrological induced pore pressure changes are separately deduced
from fifteen events when the water level exceeded six meters. The
strain measured from the extensometer show a linear contraction
during the water recharge and a nonlinear exponential extension
releasing during the water discharge. The sensitivity and stability
of the sensor are constrained by comparing continuously observed
tidal strain waves with a theoretical model. Finally, a local fault
deformation rate around $0.03 \pm 0.002$mm/yr is estimated from more
than four years' records. \clearpage
%
\section{Introduction}

The Rochefort karstic cave is located at a relatively stable
continental region where the seismicity and the risk of seismic
hazard are low. Palaeoseismology studies indicate that a maximum
$0.3$mm/yr deformation rate could account for coseismic effects from
three large earthquakes during the Pleistocene and Holocene
\cite{Camelbeeck98}. However, the recently growing stalagmite and
falling rocks inside the cave, strongly evidence that the identified
faults are active and expending in the NW-SE direction which are
driven by the regional tectonic force \cite{Vandycke01}. But the
continuous GPS measurements show motions less than $1$mm/yr which is
still inside the error bar.  Hence, to constrain the results
provided by the geological investigations, an in-situ strain
measurement experiment was conducted since 1999.

Beyond the interesting about the local faults activities, the
experiment is also focusing on providing information about the
hydrological impact on the formations process of micro fissure
networks and karstification.  There are two kinds of water flux
which contribute to the process, the slow seepage and fast flow
drainage. To monitor the strain inside a cave can help improving the
understanding the of the karst structure, especially it's porosity
and fissure networks \cite{Genty98}.

Six extensometers were set up near two identified faults inside the
Rochefort cave, two aluminum extensometers No.1 (E1) and No.2 (E2)
in 1997 and four Pyrex aluminum extensometers No.3 (E3), No.4 (E4),
No.5 (E5), and No.6 (E6) in 2000.  The hydrological strain forces,
local faults activity, and the secular earth strain were detected
from the recording of the extensometers.

\section{Experiment Site}

\begin{figure}
\noindent\includegraphics[width=30pc]{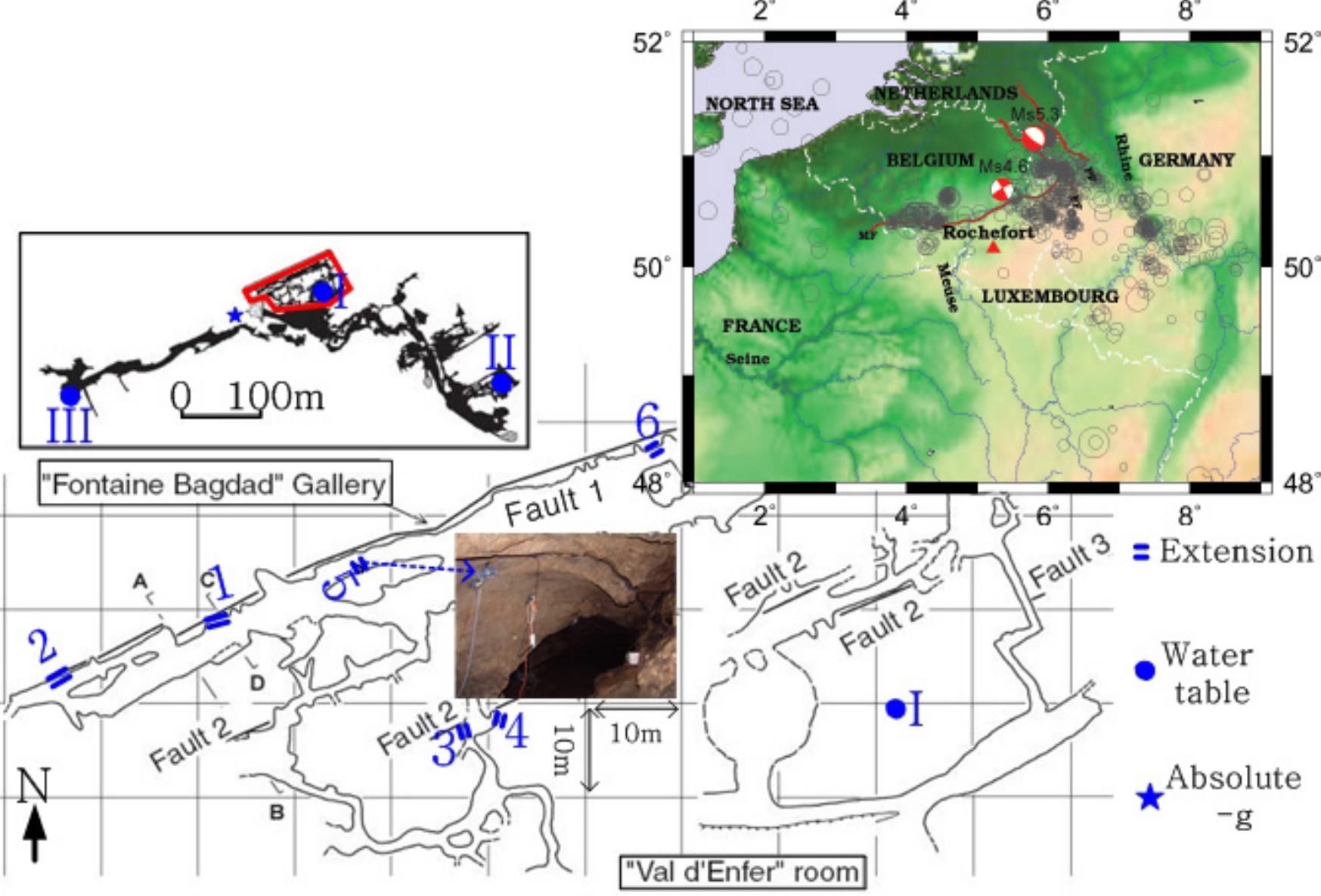} \caption{The
sketch plot of Rochefort karstic caves, the red box marks the area
at which the extensometers were installed. Three type's measurements
have been conducted: extension, water table and absolute gravity.
The maps on the upper right show the geological background of the
site. The small photo at the center is the place where E5 was
installed.} \label{Fig.Figure1}
\end{figure}

The Rochefort karstic caves are located $20$km away from the
midi-fault, along which an Ms$4.6$ earthquake occurred on Aug. 08,
1988. At the far north-east direction, this area is a relatively
active zone: all of the 'Roer Valley' graben is bounded by two
normal faults, the Peel Fault (PF) and the Feldbiss Fault (FF). An
Ms $5.3$ earthquake occurred on Apr. 13, 1992 along the PF
\cite{Camelbeeck96}. At this region, the majority earthquakes were
located at both sides of the midi-fault according to the Earthquake
Catalog of Royal Observatory of Belgium (ROB) since 1910
(Fig.\ref{Fig.Figure1}). The Rochefort cave is buried $50$ meters
underground and constituted by two galleries: large one along the
stratigraphic direction (N070E) named 'Fontaine Bagdad' and smaller
one in the dip direction 'Val d'Enfer'. Inside the cave, two types
of faults are recognized (Fig.\ref{Fig.Figure1}). The first type of
fault is contemporaneous of the Variscan folding, characterized by
reverse motion along the bedding planes. The second type is the
present active faults. The recent faults are related with
present-day tectonic activity which is evidenced by seismicity of
the neighbor area. The stereographic projection of recent faults
affecting Rochefort karstic network shows a principal NW-SE
extension is nearly perpendicular to that of the present regional
stress as illustrated by the analysis of the last strong regional
earthquake \cite{Vandycke01}.

Three water table probes have been successively installed inside the
Rochefort  karstic caves since 2005 (Fig.\ref{Fig.Figure1}). The
first one (WI) was set up under the 'Val d'Enfer' room on March
2005. The site was at the bottom of an irregular shaft with about 77
meters deep and a few meters in section. The second one (WII) was
installed at the 'Rivire des Touristes' in 2006. The 'Rivire des
Touristes' is at one of the deepest points inside the cave. It is
behaving as a river; the water comes from a sump and then flow out
into another sump. During the flood period, the nearby conduits were
filled and then the river becomes a lake. The third one (WIII) was
placed at the 'Petit Noir' on 2007, located at a large funnel shape
hall which is the ultimate point of the cave.

\section{Extensometer}

\begin{figure}
\noindent\includegraphics[width=30pc]{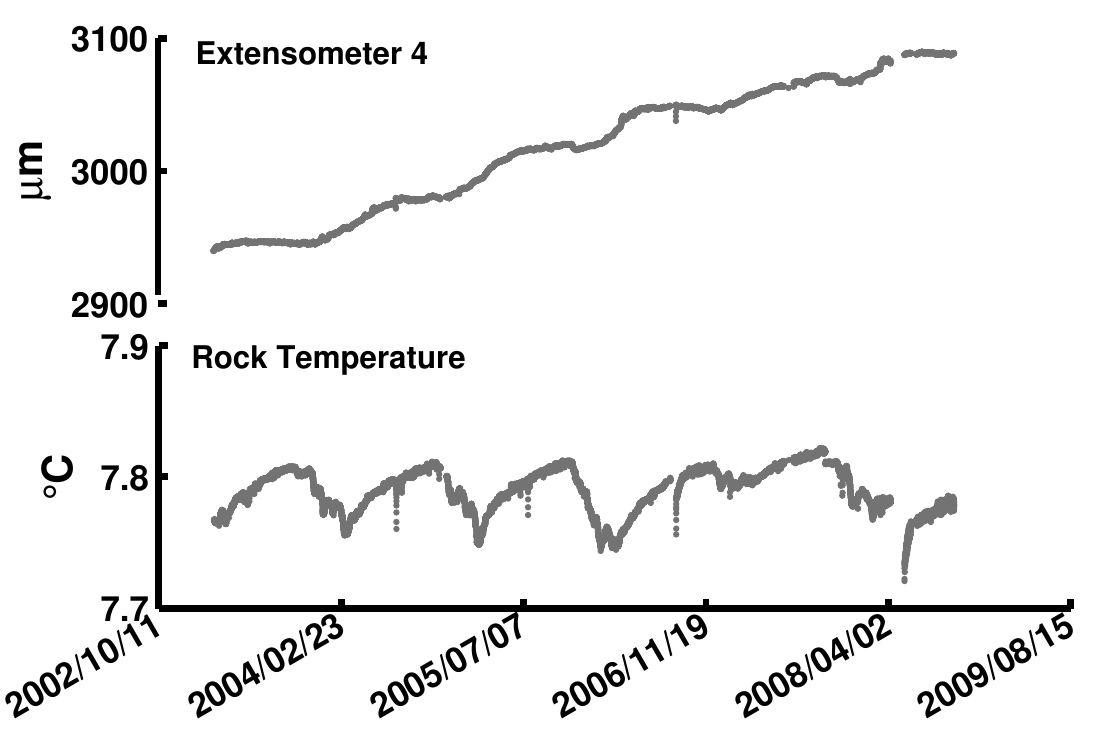}
\noindent\includegraphics[width=30pc]{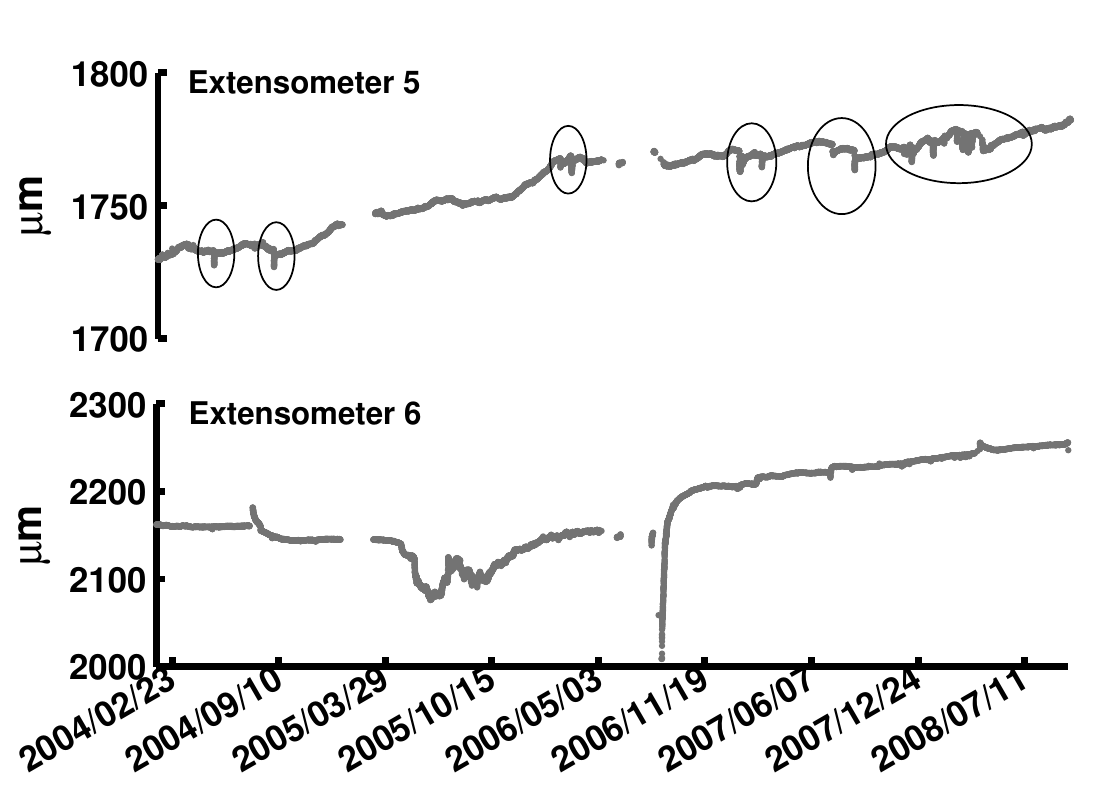} \caption{The
original data were decimated into hour sampling rate. The annual
periodical oscillations are observable from the E4 and Rock
temperature. Six ellipsoidal circles mark the hydraulically induced
deformation on E5. The pore pressure change during the dry season
induced big gaps on E6.} \label{Fig.Figure2}
\end{figure}

The Pyrex aluminum capacitive extensometers were designed and
developed in 1999 at the Royal Observatory of Belgium in Brussels.
The gauge is consisted of two round aluminum plates with $40$mm
diameter and it is fixed in the rock by two Pyrex rods. The
extensometers were installed in two steps: first, two fiducially
points are drilled $6$cm into the limestone at two beds of a fault
then the Pyrex stick are bonded to the hole by a chemical anchor
capsules (Upat-UKA3); Second, about one month later, when the piers
are adequately bounded with the rocks, we mount the extensometers
with a $1300 ^\circ$C flame. The dielectricity of the capacitance is
directly counted by an electric oscillator chip. EDAS data logger
was used as acquisition systems \cite{edas01, Francis03}. The
initial sampling interval is one minute. The laboratory calibration
experiments show that the counting frequency is linearly
proportional to the displacement when the relative displacement is
less than $100\mu$m.  The stability and sensitivity of the gauges
are influenced by the humidity, pressure, and temperature since the
sensors are completely exposed to the environment. Fortunately, the
meteorological conditions are very stable inside the cave. The
continuous rock temperature monitoring near E4 gave an annual
temperature variations of about $0.06 ^ \circ$C, and the maximum
daily changes were less than $0.02 ^ \circ$C. The thermal expansion
coefficient is $22.3 \times 10 ^{-6}$ for aluminum and $3.25 \times
10 ^{-6}$ for Pyrex glass. Thus, taking into account the laboratory
tests, the expected strain resolution should be better than 0.1ppm.
The original data were decimated into hour sampling rate
(Fig.\ref{Fig.Figure2}). We selected the most homogeneous records of
E4 (length $81$mm, azimuth $84^\circ$), E5 (length $92$mm, azimuth
$136^\circ$) and E6 (length $75$mm, azimuth $58^\circ$) since 2003.
Three years after the installation, therefore we can believe that
the extensometers are adequately coupled with the host rocks.

\section{Results}
\subsection{Tidal Strain}

The most difficult problem for such experiment is the calibration
and evaluation of the stability of the sensors. Since the solid
Earth Tides are continuously registering on the E4, E5, and E6,
which provide us a stable reference. The tidal strain measurement
began from 1951 when Sassa, Ozawa and Yoshikawa published the first
results obtained with a superinvar wire extensometer
\cite{Melchior83}. After that, a great effort have been paid to
design high accuracy strainmeters by different groups, among them
the most reliable gauges should be the Laser strain meter
\cite{Berger70, King73,Berger73,Sydenham74,Beavan77}. With the new
technology, the secular earth strain tides can be easily observed in
different places. Thus, the interests of strain observations have
been moved from the earth tide to seismic toroidal modes, tectonic
activities, hydrological effects and transient pre-, co-, or post-
seismic signals \cite{Zadro99}.

\begin{figure}
\noindent\includegraphics[width=30pc]{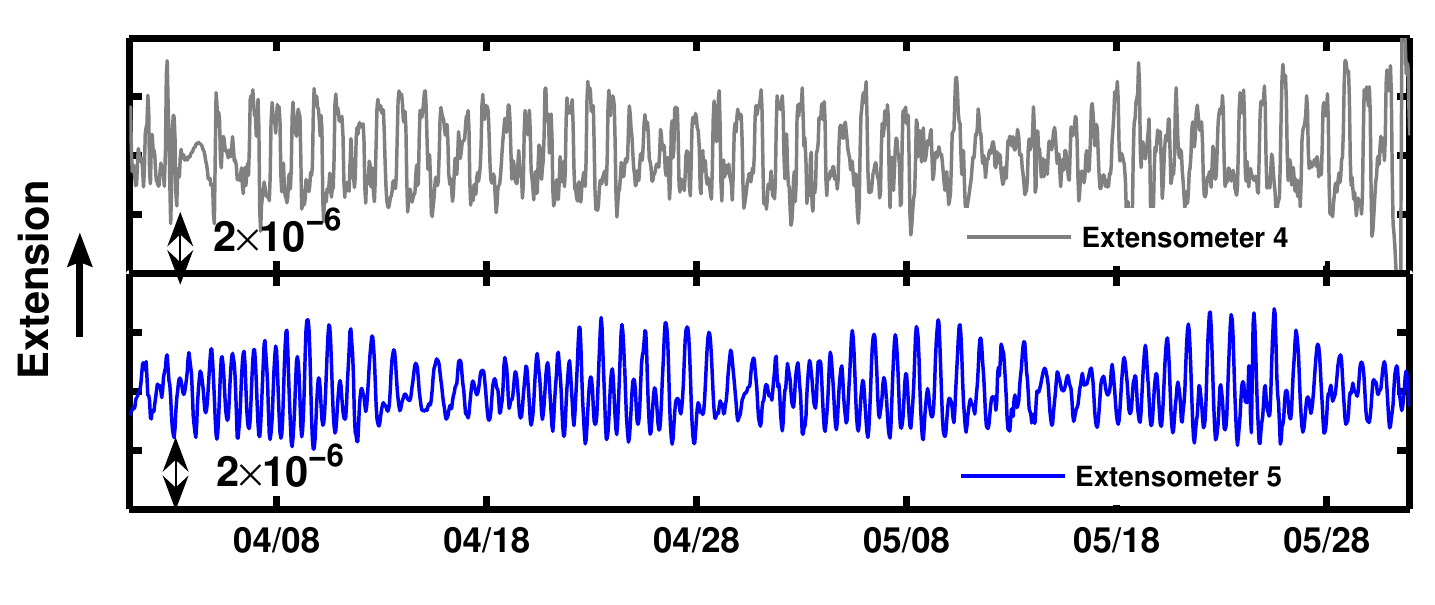}
\noindent\includegraphics[width=28pc]{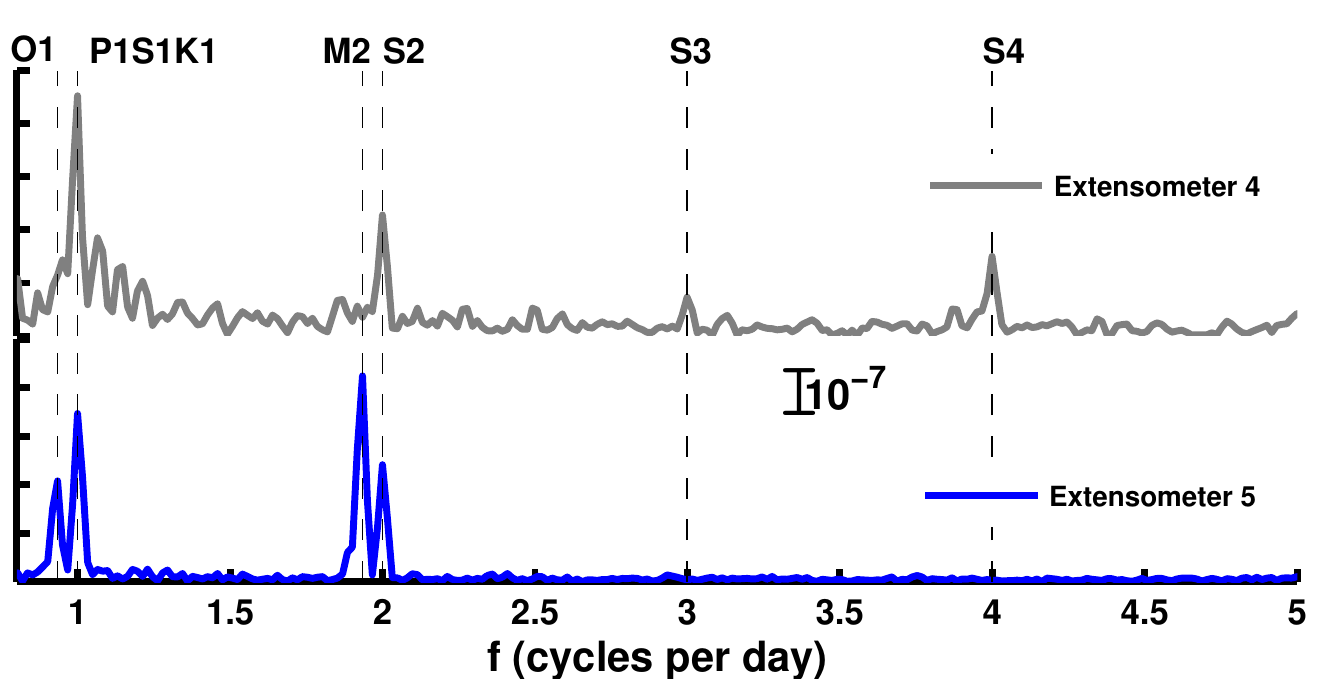}
\caption{(Upper), The original data have been band pass filtered
between 0.6 cpd and 8 cpd, the earth tides are obviously recorded by
two gauges, (Lower), Amplitudes Fourier Spectrum of extensomter No.
4 and No.5.} \label{Fig.Figure3}
\end{figure}

\begin{figure}
\noindent\includegraphics[width=10pc]{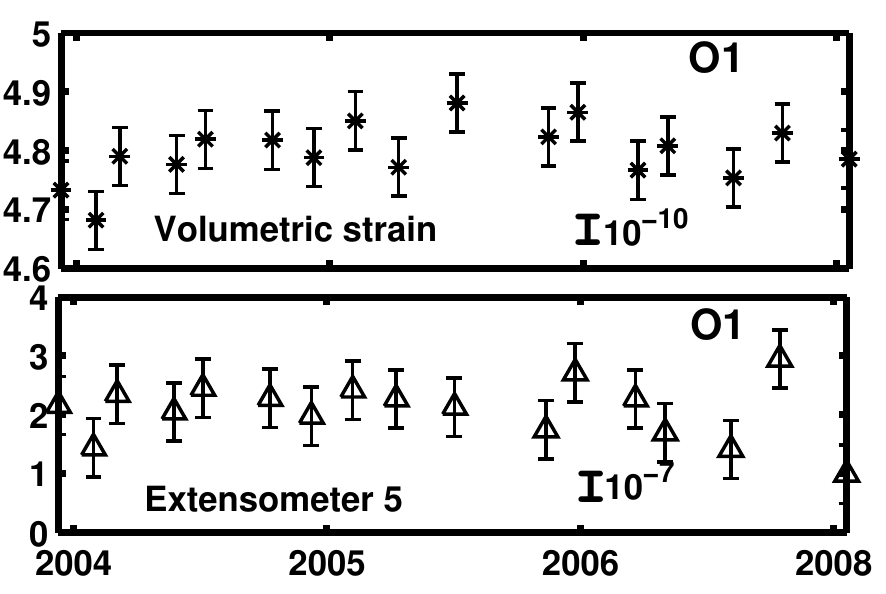}
\noindent\includegraphics[width=10pc]{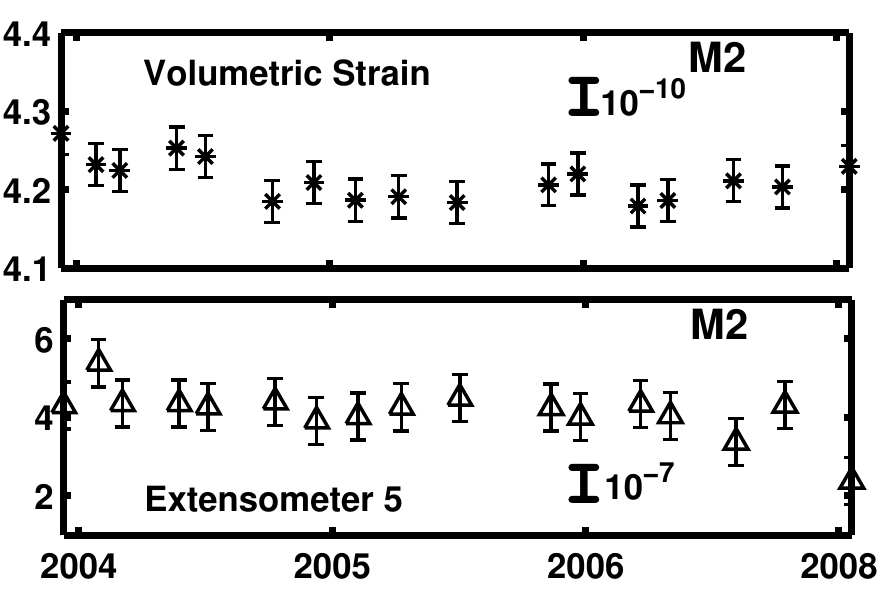}
\noindent\includegraphics[width=12pc]{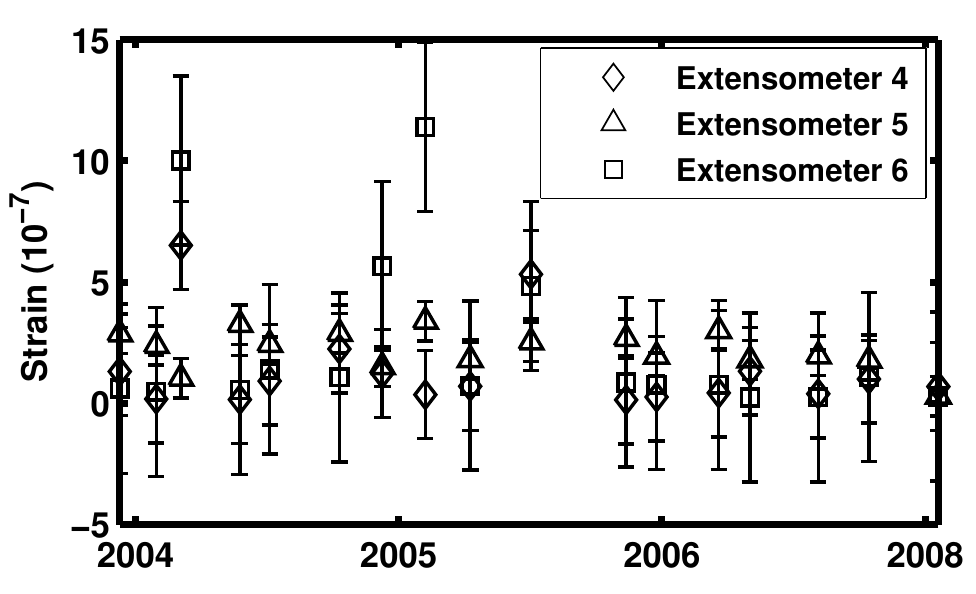} \caption{(a) O1
amplitudes distribution upper channel synthetic volumetric tidal
strain, lower channel observed O1 tides from E5. (b) M2 amplitudes
distribution upper channel synthetic volumetric tidal strain, lower
channel observed M2 tides from E5. (c) S2 components separated from
three sensors E4, E5 and E6.} \label{Fig.Figure4}
\end{figure}

Due to the high accuracy of the tidal model, the tides are valuable
input signals for the instruments. It can be applied to calibrate
and check the sensitivity of any arbitrary records in situ such as
the borehole strain measurements \cite{zurn01}. The diurnal and
semi-diurnal components have been founded from the in situ relative
seismic velocity variation records in Japan \cite{Yamamura03}, by
which the relation of relative seismic velocity variation with
strain can be directly measured. The minimum resolution of our
capacitive extensometer can reach $10^{-7}$. The synthetic
volumetric earth strain tides computed with HW95 tidal potential
\cite{Eterna96} gave the amplitudes are between $10^{-8}$ and
$10^{-9}$ at the surface of Rochefort karstic caves. Thus the gauges
can hardly record the tidal strain by principle. But it is
undoubtedly registered at the extensometer 4 and 5. To check the
tidal frequency band signals, the data were band pass filtered
between $0.6$ cycles per day and $8$ cycles per day. Two month's
results between April, 1st and May 31st in 2005 were plotted
(Fig.\ref{Fig.Figure3} upper). The Amplitude Fourier Spectrum shows
the diurnal and semidiurnal solar principal waves and its harmonic
components until the fourth (Fig.\ref{Fig.Figure3} lower). These
signals are dominated at the E4 which could be the thermal induced
elastic deformation. Similar phenomenon has been reported from a
long base line laser strain meter measurements \cite{Berger73}. The
principal lunar diurnal wave O1 and semidiurnal wave M2 are
permanently registered in the gauges E5 with unexpected high
signal-to-noise ratio. The theoretical values of the volumetric
strain in the azimuth of E5 are computed using Eterna3.3
\cite{Eterna96} from Apr.01, 2005 to Sep.30, 2005. We compared the
phases of O1, M2 and S2 with the records of E5
(Fig.\ref{Fig.Figure4}). The observed phases (O1, M2 and S2) were
close to the values of the hydrostatical tidal strain (Table
\ref{tab:table1}). It evidences that the sensors are recording the
hydrostatic strains induced by pore pressure changes but the
amplitudes were two orders larger than the synthetic values.

\subsection{ Hydrological Effects }

\begin{figure}
\noindent\includegraphics[width=20pc]{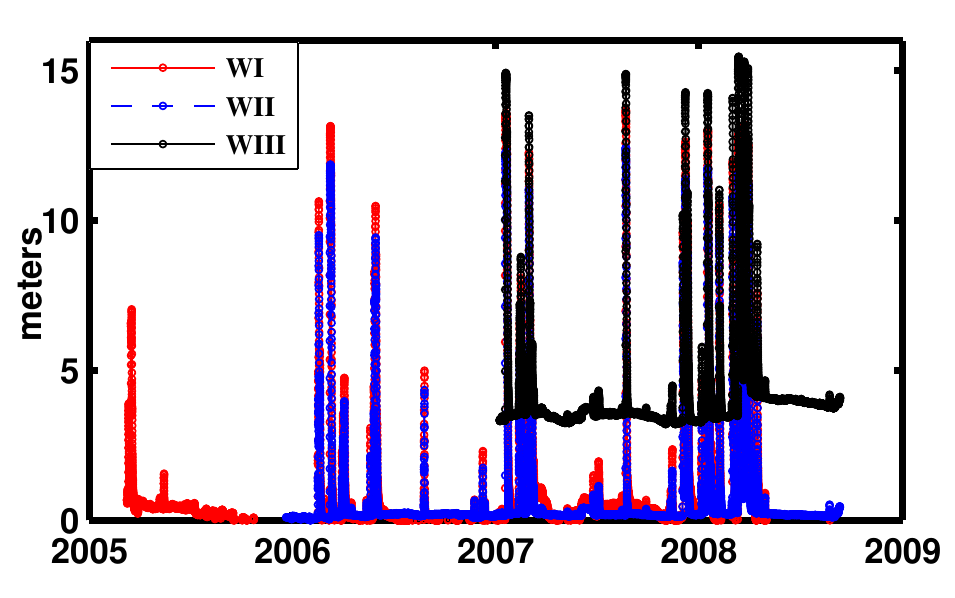}
\noindent\includegraphics[width=20pc]{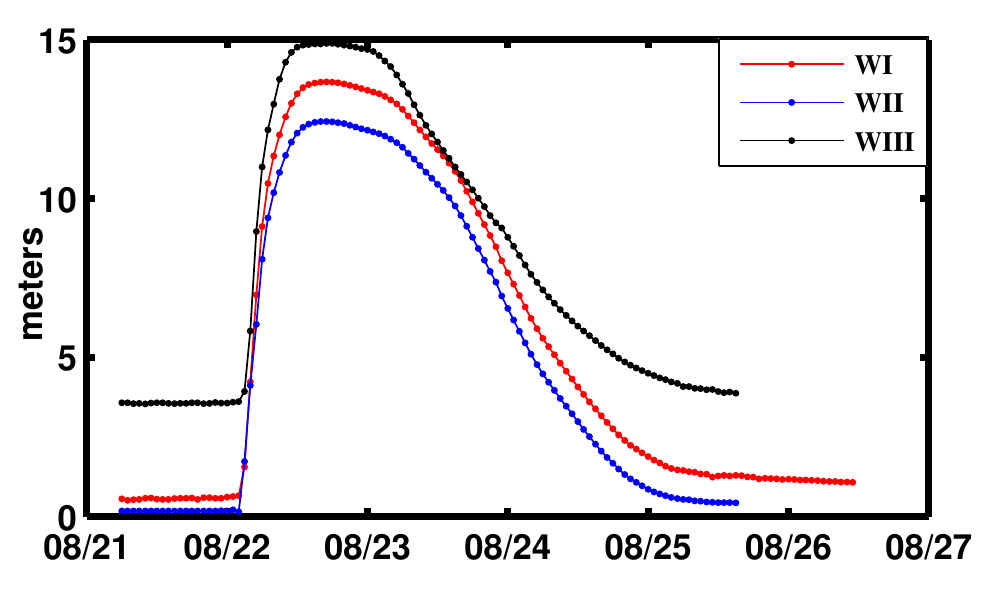} \caption{(Left),
water table records at three locations of the karstic caves, there
are 16 times that the water level rose higher than 6 meters.
(Right), a typical flood event recorded by three water tables.}
\label{Fig.Figure5}
\end{figure}

\begin{figure}
\noindent\includegraphics[width=30pc]{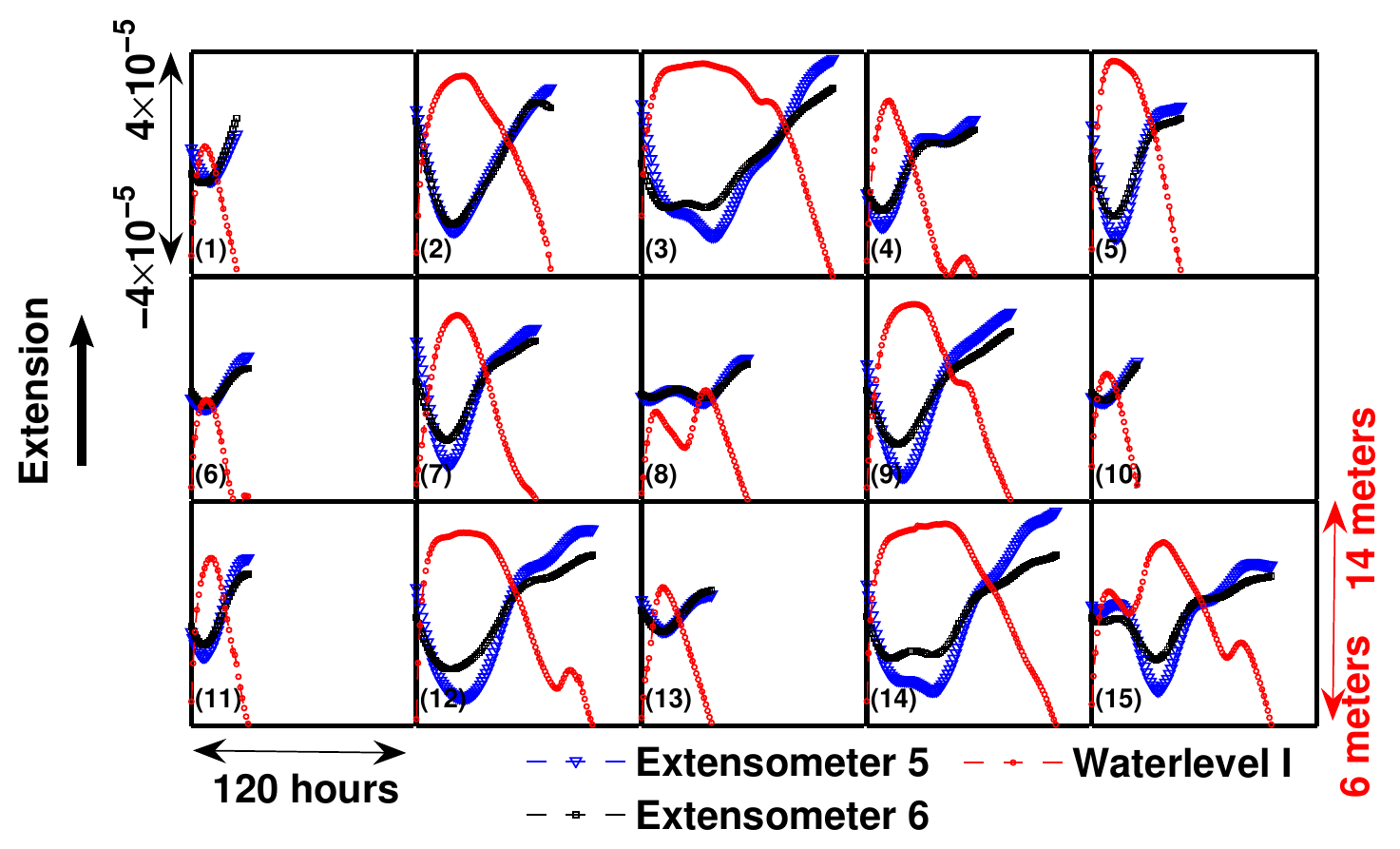} \caption{The
water table WI, Extensometer 5, and 6 are plotted together when the
water level is higher than 6 meters. The number of each box
represents the hydrological event from 2004 to 2008.}
\label{Fig.Figure6}
\end{figure}

The experiment site suffered a relatively dry season from the Feb,
2005 to the beginning of 2006, and then the precipitations were
increased. It becomes a rainy season from the mid of 2007 to the end
of 2008 (Fig.\ref{Fig.Figure5} left). When the Lomme River was
flooded, the water level rose and fell simultaneously at WI, WII and
WIII. The shortage differences make three aquifers discharging water
with slightly varied speeds (Fig.\ref{Fig.Figure5} right). The WI
water level gauge is the nearest probe to the extensometers so that
the hydrological induced strain was directly measured on E5 and E6
from WI. To measure the strain changes from the extensometers, a
hydrological event is defined when the water level reaches higher
than 6 meters. We choice such a definition because the hydrological
strain forces signal was superposed on a long term slop of fault
motion.  Sixteen events were located from the records of the water
probe since 2005 until 2008. Due to the failure of the power supply,
the E5 and E6 have missed one event. The E5 and E6 are well
correlated with the processing of water charging and discharging at
WI site (Fig.\ref{Fig.Figure6}). Obviously, the response of the
extensometers to the charging and discharging procedure were
different. The pore pressure was increased when the water level rose
and decreased with water discharge.

The fifteen isolated hydrological events were divided into three
categories according to the duration of the highest water level.
Type one, the process of water charging and discharging is shorter
than forty hours. The second type is longer than forty hours, and
the third type, there is more than once that the height of water
table extended $6$ meters in a $150$ hours' window (Table
\ref{tab:table2}). The highest water level was set as the boundary
point between the water charging and discharging.  We separately
measured the strain changes as a function of water level. Six events
were fallen into the first category, seven events belong to the
second category, and two were in the last category. During the water
charging, the strain rate was linearly proportional to the water
level and flux.

\begin{figure}
\noindent\includegraphics[width=11pc]{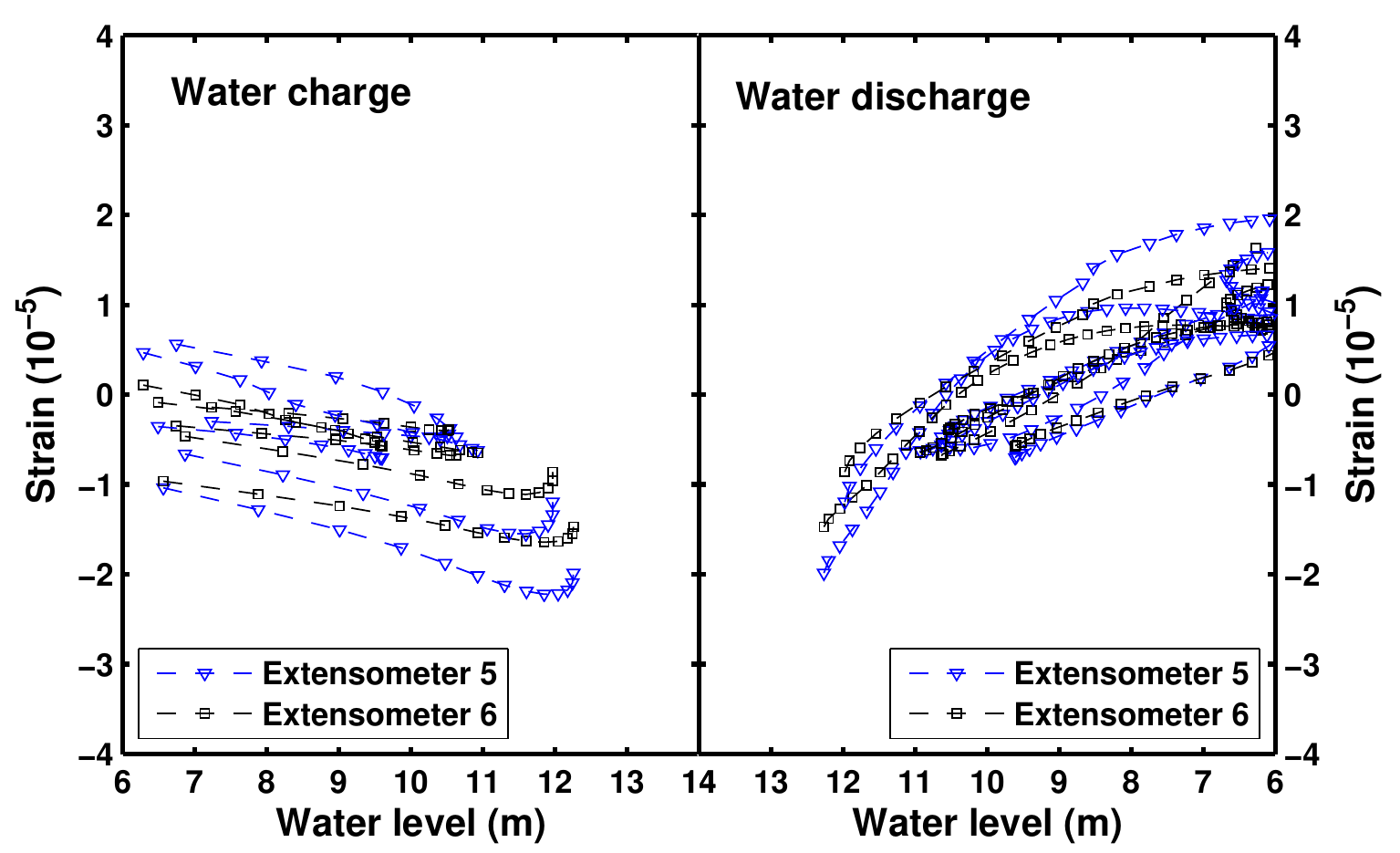}
\noindent\includegraphics[width=11pc]{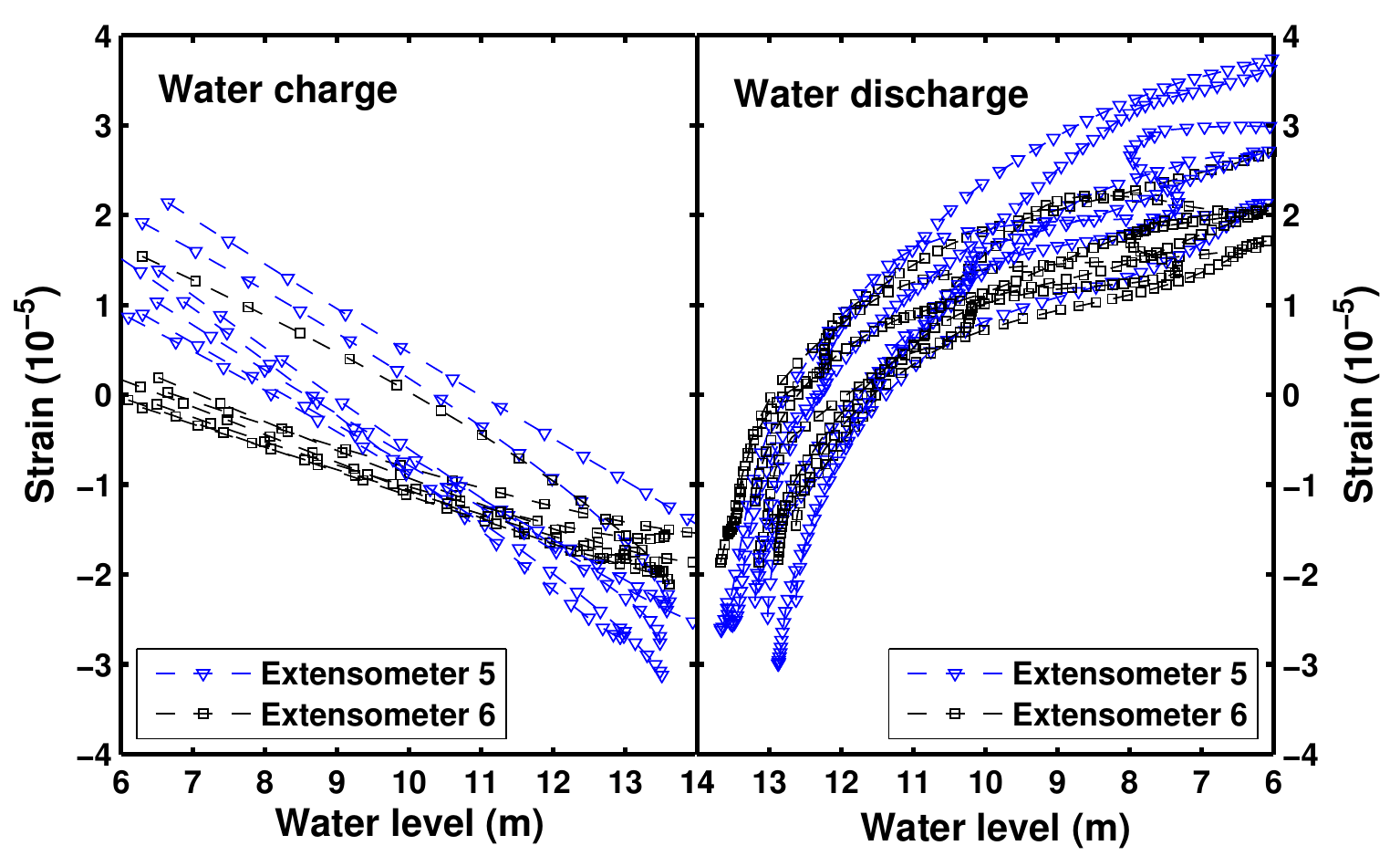}
\noindent\includegraphics[width=11pc]{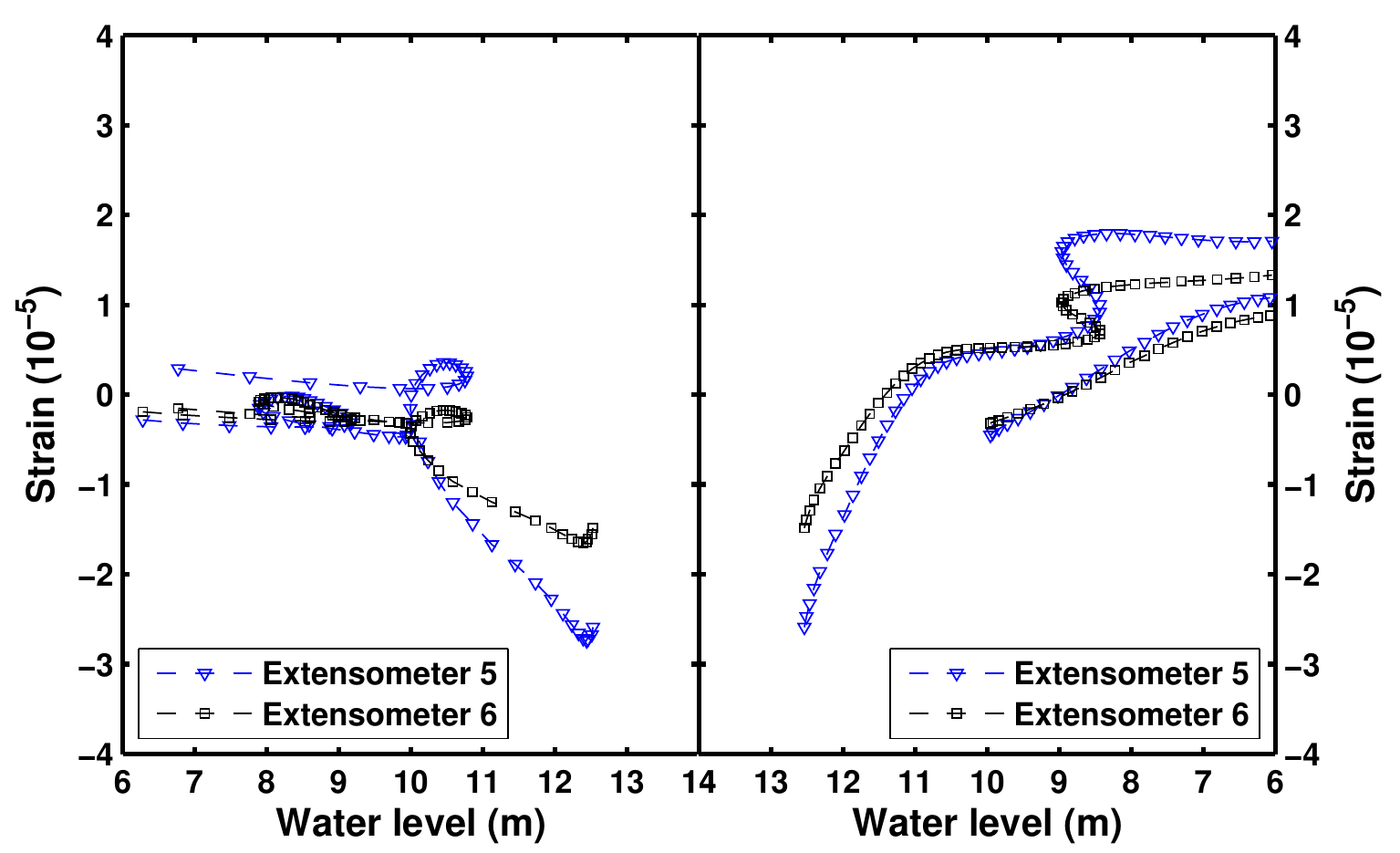} \caption{The
strain versus water level. (a) The first type, the duration of water
level higher than 6 meters is shorter than 10 hours. (b) The second
type, the duration of water level higher than 6 meters is longer
than 10 hours. (c) The third type, more than one times the water
level rose higher than 6 meters.} \label{Fig.Figure7}
\end{figure}

At the stage of the water discharging, the strain rate are nonlinear
exponentially recovery to the water level (Fig.\ref{Fig.Figure7}a).
It evidently follows this pattern for all events of the second type
(Fig.\ref{Fig.Figure7}b). The eighth and fifteenth events belong to
the third type, the karst fissure networks were charging and
discharging more than one times in a $150$ hours window. Both
extensometers were instantaneously responding to about one meters
water level rising and falling (Fig.\ref{Fig.Figure7}c).

The extensometers simultaneously react to the pore pressure changes
which suggest that the pore pressure was altered mainly due to the
karst fissure networks were rapidly filling and run-off during the
rainy season. In the regional scale, it shows that the complete
karstic caves work as an elasticity and homogeneousness media in a
short term.

\subsection{ The fault activities inside the karstic cave }

\begin{figure}
\noindent\includegraphics[width=30pc]{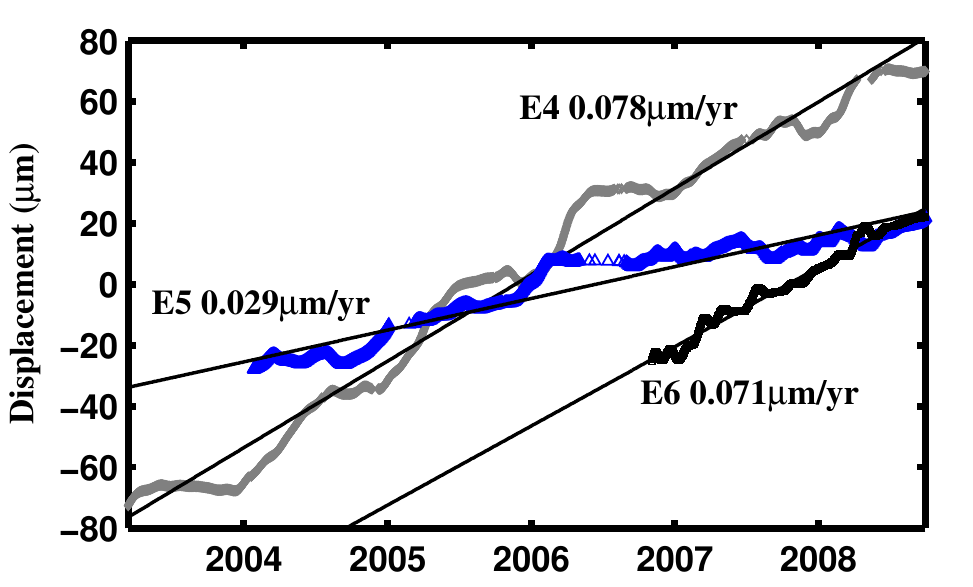} \caption{The long
term trend of E4, E5 and E6 is fitted by a linear function with
95$\%$ confidential level. } \label{Fig.Figure8}
\end{figure}

From the beginning of 2005 to the February of 2006, it was a
relatively dry season. The water content was decreased inside the
cave which altered hydrostatic strain inside the cave. The abnormal
contraction appeared on the E6 can be counted for such effects
(Fig.\ref{Fig.Figure2}). In 2006, the E5 and E6 were interrupted by
a failure of the power supply. The E6 restarted to work after the
power supply was repaired. The records of E4, E5, and E6 are
decimated into daily sampling rate and low pass filtered to only
keep the long term trends with frequency lower than one cycle per
month (Fig.\ref{Fig.Figure8}). The annual rock temperature change is
$0.06 ^ \circ$ near the E4 by which the annual thermal induced
elastic deformation is deduced from the E4 with amplitude $7.3 \pm
1.9 \mu$m. The thermal effect is obviously weaker at the site of
extensometer 5 and 6 where the water are penetrating through the
karstic fissure networks and some new stalactites are growing
nearby. Each of these three long term trenches was separately fitted
by a linear function with $95\%$ confidential boundary. The E4 shows
a deformation rate of $0.028$mm/yr and $0.026$mm/yr for E6. The E6
is bonding across the fault 1 and E4 is to the fault2. It is
reasonable to find that the E5 shows a slower expanding rate
$0.010$mm/yr than the E4 and E6 because the E5 is not directly
attached to the identified active faults. The extension direction of
the fissure where E5 installed, are nearly perpendicular to the
fault 1 and 2.

\section{Discussion and conclusions}

Diurnal and semidiurnal tidal constituents are continuously recorded
by the E5. However, the amplitudes are much greater than the
synthetic earth strain tides. But the observed tendency of O1
amplitudes distribution agree well with theoretical value which
provides valuable constrains about the stability of the gauge. The
solar principal wave S2 and its harmonic components are permanently
recorded by the E4 and sometimes the E6. The semidiurnal S2 wave
induced by thermal-elastic oscillation with amplitude around $2.4
\times 10^{-7}$ is clearly observable on the E4. By inter comparing
the S2 waves among three sensors, the S2 amplitudes differences were
less than $10 ^ {-7}$, which confirmed the sensor's stability can
reach $0.1$ppm (Fig. 4). The amplitudes of observed tidal strain
waves are much greater than the theoretical value which could be
attributed to the cavity effects, but due to the chaos of the
fractal cavities, it is difficult to model such effect.

The hydrological agents induced rock deformations were observed from
the E5 and E6. The water conductivity and porosity altered the pore
pressure inside the cave. These hydrological induced deformations
have been found in other geodetic observations. Several solutions
are proposed to correct it from the geodetic observations, among
them the mostly used methods are the rain function \cite{Langbein90}
and the predictive filtering methods \cite{Braitenberg99}. The E5
and E6 have different maximum strain response to water level. But
the rate of strain accumulation and release accompanied by the
processing of water charging and discharging, gave coherent results
by both gauges for all fifteen events. During the water charging, it
shows a linearly contraction at E5 and E6 induced by the buoyancy
force, at the period of the water discharging, the strain is
recovering with an exponential trends due to the pore pressure was
altered by the conductivity and porosity of the media during the
flood periods. The contraction rate induced by the water level
rising of E5 is slightly faster than E6, the extension after the
water releasing, E5 and E6 shared similar rate.

The absolute gravity measurement at the site, shows gravity value
increases $90$nm/s$^2$ when the water level rose more than $10$
meters \cite{mvc06}. The gravity changes calculated from a $10$
meters thick cylinder model in $60$ meters depth and with $5\%$
porosity, is $250$nm/s$^2$ Dr. M.Van Camp (pers. comm.). The
observed gravity changes are about $1/3$ of the modeled value. Since
the strain were linearly reacted to water charging, it suggested
that both the strain and gravity change, can be attributed to the
buoyancy force during the water filling the karst fissure networks.

The E6 showed a $0.026$mm/yr displacement of fault 1 and E4 gave a
$0.028$mm/yr displacement of fault2 which was superposed with a $7.3
\pm 1.9 \mu$m annual oscillation term.  A perpendicular fissure to
the fault 1 measured by the E5, gave a $0.010$mm/yr rate. The
fissure is $20$cm in width so that the formation of this fissure is
about $20, 000$ years if we suppose that the activity of the region
is relative stable since it. The expending rate deduced from the E4
and E6, agreed with the results obtained by stereographic projection
\cite{Vandycke01}. It suggests that a local background driving
displacement rate is about $0.03 \pm  0.002 $mm/yr.

A factual karstic cave is normally thought as an unfavorable site
for the instrumental observations due to its complex geometry with
strong heterogeneity and high humidity rates. But water drips
monitoring under the stalactites, have revolved broad interesting
for the karstification processing and palaeoclimatology study
\cite{Genty08}. The preliminary results of our strain measurement
shed new light on the geophysical experiments, the result suggest
that if the sensor is properly installed,  maximum following the
characteristics of the nature, the ability of detecting very weak
signals like the hydrology, local fault activities  and secular
earth strain can be achieved.

\section*{Acknowledgments}
We are very grateful to the support of Dr. Ronald Van der Linden,
director general of Royal Observatory of Belgium. We benefit a lot
from the discussions with Dr. Michel Van Camp. We used the ROB
earthquake catalog. The first author is financially supported by the
Action 2 contract from the Belgian Ministry of Scientific Politics.
The experiments in Rochefort were supported by the Ministry of the
Wallon Region(Belgium).

\bibliographystyle{Jcks}

\begin{thebibliography}{20}
\expandafter\ifx\csname
natexlab\endcsname\relax\def\natexlab#1{#1}\fi
\expandafter\ifx\csname url\endcsname\relax
  \def\url#1{\texttt{#1}}\fi
\expandafter\ifx\csname urlprefix\endcsname\relax\def\urlprefix{URL
}\fi

\bibitem[{Beavan and Goulty(1977)}]{Beavan77}
Beavan, R., Goulty, N., 1977. Earth-strain observations made with
the
  \textsc{C}ambridge laser strainmeter. Geophys.J.R.Astr.Soc. 48, 293--305.

\bibitem[{Berger and Lovberg(1970)}]{Berger70}
Berger, J., Lovberg, R., 1970. Earth strain measurements with a
laser
  interferometer. Science 170, 296--303.

\bibitem[{Berger and Wyatt(1973)}]{Berger73}
Berger, J., Wyatt, F., 1973. Some observations of \textsc{E}arth
strain tides
  in \textsc{C}alifornia. Phil.Trans.R.Soc.Lond.A. 274, 267--277.

\bibitem[{Braitenberg(1999)}]{Braitenberg99}
Braitenberg, C., 1999. Estimating the hydrologic induced signal in
geodetic
  measurements with predictive filtering methods. Geophys.Res.Lett. 26,
  775--778.

\bibitem[{Camelbeeck and Meghraoui(1996)}]{Camelbeeck96}
Camelbeeck, T., Meghraoui, M., 1996. Large earthquakes in northern
  \textsc{E}urope more likely than once thought. EOS Trans., Am. Geophys. Union
  77, 405--409.

\bibitem[{Camelbeeck and Meghraoui(1998)}]{Camelbeeck98}
Camelbeeck, T., Meghraoui, M., 1998. Geological and geophysical
evidence for
  large palaeoearthquakes with surface faulting in the \textsc{R}oger
  \textsc{G}raben (\textsc{N}orthwesten \textsc{E}urope). Geophys.J.Int. 132,
  347--362.

\bibitem[{Genty(2008)}]{Genty08}
Genty, D., 2008. Palaeoclimate research in \textsc{V}illars
\textsc{C}ave
  (\textsc{D}ordogne, \textsc{SW}-\textsc{F}rance). International Journal of
  Speleology 37(3), 171--191.

\bibitem[{Genty and Deflandre(1998)}]{Genty98}
Genty, D., Deflandre, G., 1998. Drip flow variations under a
stalactite of the
  \textsc{P}\`{e}re \textsc{N}o\"{e}l cave (\textsc{B}elgium).
  \textsc{E}vidence of seasonal variations and air pressure constraints.
  Journal of Hydrology 211, 208--232.

\bibitem[{King and Bilham(1973)}]{King73}
King, G., Bilham, R., 1973. Tidal tilt measurements in
\textsc{E}urope. Nature
  243, 74--75.

\bibitem[{Langbein et~al.(1990)Langbein, Burford, and Slater}]{Langbein90}
Langbein, J., Burford, R., Slater, L., 1990. Variations in fault
slip and
  strain accumulation at \textsc{P}arkfield, \textsc{C}alifornia:initial
  results using two color geodimeter measurements. J.Geophys.Res. 95,
  2533--2552.

\bibitem[{Melchior(1983)}]{Melchior83}
Melchior, P., 1983. The Tides of the Planet Earth, Pergamon Press.
New York.

\bibitem[{Sondag et~al.(2003)Sondag, \textsc{v}an Ruymbeke, Soubi\'{e}s.F.,
  Santos, Somerhausen, Seidel, and Boggiani}]{Francis03}
Sondag, F., \textsc{v}an Ruymbeke, M., Soubi\'{e}s.F., Santos, R.,
Somerhausen,
  A., Seidel, A., Boggiani, P., 2003. Monitoring present day climatic
  conditions in tropical caves using an environmental data acquisition system
  \textsc{(EDAS)}. Journal of Hydrology 273, 103--118.

\bibitem[{Sydenham(1974)}]{Sydenham74}
Sydenham, P., 1974. Where is experimental research on earth strain?
Nature 252,
  278--280.

\bibitem[{Van~Camp et~al.(2006)Van~Camp, Meus, Quinif, Kaufmann, \textsc{v}an
  Ruymbeke, Vandiepenbeeck, and Camelbeeck}]{mvc06}
Van~Camp, M., Meus, P., Quinif, Y., Kaufmann, O., \textsc{v}an
Ruymbeke, M.,
  Vandiepenbeeck, M., Camelbeeck, T., 2006. Karst aquifer investigation using
  absolute gravity. EOS Trans., Am. Geophys. Union 87, 298--299.

\bibitem[{Van~Ruymbeke et~al.(2001)Van~Ruymbeke, Beauducel, and
  Somerhausen}]{edas01}
Van~Ruymbeke, M., Beauducel, F., Somerhausen, A., 2001. The
environmental data
  acquisition system \textsc{(EDAS)} developped at the \textsc{R}oyal
  \textsc{o}bservatory of \textsc{B}elgium. Journal of the Geodetic Society of
  Japan 47, 40--46.

\bibitem[{Vandycke and Quinif(2001)}]{Vandycke01}
Vandycke, S., Quinif, Y., 2001. Recent active faults in
\textsc{B}elgian
  \textsc{A}rdenne revealed in \textsc{R}ochefort \textsc{K}arstic network
  (\textsc{N}amur province, \textsc{B}elgium). Netherlands Journal of
  Geosciences 80, 297--304.

\bibitem[{Wenzel(1996)}]{Eterna96}
Wenzel, H.-G., 1996. The nanogal software: Earth tide data
processing package
  eterna 3.3. Bulletin d'Informations Mar\'{e}es Terrestres 124, 9425--9439.

\bibitem[{Westerhaus and Z\"{u}rn(2001)}]{zurn01}
Westerhaus, M., Z\"{u}rn, W., 2001. On the use of earth tides in
geodynamic
  research. Journal of the Geodetic Society of Japan 47 No.1, 1--9.

\bibitem[{Yamamura et~al.(2003)Yamamura, Sano, Utada, Takei, and
  Nakao}]{Yamamura03}
Yamamura, K., Sano, O., Utada, H., Takei, Y., Nakao, S., 2003.
Long-term
  observation of in situ seismic velocity and attenuation. J.Geophys.Res. 108,
  doi:10.1029/2002JB002005.

\bibitem[{Zadro and Braitenberg(1999)}]{Zadro99}
Zadro, M., Braitenberg, C., 1999. Measurements and interpretations
of
  tilt-strain gauges in seismically active areas. Earth-Science Reviews 47,
  151--187.

\end{thebibliography}

\clearpage
%
%
%
%

\clearpage

%
%
\begin{table}
\begin{threeparttable}
\singlespace \caption{Tidal analysis result of E5 records between
Apr.01 and Sept.30 2005.}
\begin{tabular}{cccrrrr} \toprule
      Wave & E5($10^{-7}$) & Syn($10^{-9}$) & Ratio & E5($\phi$) & Syn($\phi$) & Diff.  \\ \hline
      O1   & 2.219           & 5.006            & 44.3  & 138.311    & 125.186     & 13.125 \\
      M2   & 3.798           & 4.593            & 82.7  & 134.988    & 138.229     & -3.241 \\
      S2   & 1.859           & 1.623            & 114.5 & 12.112     & 8.122       & 3.990  \\ \bottomrule
\end{tabular}
\label{tab:table1}
\end{threeparttable}
\end{table}

\begin{table}
\begin{threeparttable}
\singlespace  \caption{Hydrological events were separated into three
groups, the maximum strain changes corresponding to the highest
water is measured from each event.}
    \begin{tabular}{rcrrcc} \toprule
    NO. &   Date            & Dur(hrs) & WI(m) & E5($10^{-5}$)  & E6($10^{-5}$)  \\ \hline
    1   & 2006/02/17 00:00  & 25       & 10.6  & -1.18          & -0.33          \\
    4   & 2007/03/01 11:00  & 5        & 12.3  & -1.20          & -0.68          \\
    6   & 2007/12/03 03:00  & 31       & 9.6   & -0.09          & -0.13          \\
    10  & 2008/02/06 09:00  & 25       & 10.5  & -0.04          & -0.09          \\
    11  & 2008/03/01 09:00  & 9        & 11.9  & -0.90          & -0.65          \\
    13  & 2008/03/16 20:00  & 38       & 10.9  & -1.11          & -0.76          \\ \hline
    2   & 2006/03/09 06:00  & 72       & 13.1  & -4.35          & -3.69          \\
    3   & 2007/01/18 06:00  & 102      & 13.6  & -4.75          & -1.56          \\
    5   & 2007/08/22 05:00  & 48       & 13.7  & -4.05          & -2.05          \\
    7   & 2007/12/06 23:00  & 64       & 12.6  & -4.42          & -2.09          \\
    9   & 2008/01/16 02:00  & 77       & 13.0  & -4.04          & -1.92          \\
    12  & 2008/03/11 08:00  & 94       & 12.9  & -3.93          & -1.84          \\
    14  & 2008/03/21 06:00  & 101      & 13.2  & -3.84          & -1.49          \\ \hline
    8   & 2007/12/10 05:00  & 57       & 9.9   & -0.19          & -0.13          \\
    15  & 2008/03/31 22:00  & 96       & 12.5  & -3.07          & -1.49          \\ \bottomrule
    \end{tabular}
\label{tab:table2}
\end{threeparttable}
\end{table}


\clearpage

%
%

%
\end{document}